\documentclass{PoS}
\pdfoutput=1 

\usepackage{graphicx}
\usepackage{xspace}
\usepackage{booktabs}
\usepackage{diagbox}

\newcommand\DDaAMG{\mbox{DDalphaAMG}\xspace}

\title{Multigrid for Wilson Clover Fermions in Grid}

\ShortTitle{Multigrid for Wilson Clover Fermions in Grid}

\author{\speaker{Daniel Richtmann}$^{\:a}$, Peter A. Boyle$^b$, Tilo Wettig$^a$\\
  Department of Physics, University of Regensburg, 93040 Regensburg, Germany\\
  School of Physics, The University of Edinburgh, Edinburgh EH9 3JZ, UK}

\abstract{
  With the ever-growing number of computing architectures, performance portability is an important aspect of (Lattice QCD) software.
  The Grid library provides a good framework for writing such code, as it thoroughly separates hardware-specific code from algorithmic functionality and already supports many modern architectures.
  We describe the implementation of a multigrid solver for Wilson clover fermions in Grid by the RQCD group.
  We present the features included in our implementation, discuss initial optimization efforts, and compare the performance with another multigrid implementation.}

\FullConference{The 36th Annual International Symposium on Lattice Field Theory - LATTICE2018\\
		22-28 July, 2018\\
		Michigan State University, East Lansing, Michigan, USA.}

\begin{document}

\section{Introduction}

State-of-the-art Lattice QCD simulations are performed with (nearly) physical quark masses on large and fine lattices.
The resulting numerical cost of inverting the Dirac matrix can be reduced significantly by multigrid (MG) methods.
Constructing a working MG for Lattice QCD turned out to be a complicated endeavor, but by now MG algorithms exists for all major fermion formulations: Wilson/clover \cite{Luscher:2007se,Brannick:2007ue,Babich:2010qb,Osborn:2010mb,Frommer:2013fsa}, domain wall \cite{Cohen:2012sh,Boyle:2014rwa,Yamaguchi:2016kop}, overlap \cite{Brannick:2014vda}, twisted mass \cite{Alexandrou:2016izb}, staggered \cite{Brower:2018ymy}.

In the typical workflow of a (Lattice QCD) programmer, there is tension between the desire for easy code development and the need to obtain high performance.
Grid \cite{Boyle:2016lbp,Boyle:github} is a data-parallel library that aims to resolve this tension.
It addresses all three major parallelization paradigms on CPUs: (i) SIMD based on site fusing, (ii) threading based on OpenMP (fork-join model), and (iii) message passing based on MPI.
Grid exposes an elegant high-level interface that allows for rapid code development.

New features are constantly being added to Grid.
The present contribution describes the implementation of an MG solver for Wilson clover fermions in Grid \cite{Richtmann:github}, including significant performance optimizations that also benefit other fermion formulations.

\section{Implementation details}
\label{sec:implementation}

\subsection{Features}

Our work is based on the MG building blocks already present in Grid, which where developed in the implementation of the HDCR algorithm for domain-wall fermions \cite{Yamaguchi:2016kop}.
In particular, two classes are of relevance:
The intergrid operators are implemented in \texttt{Aggregation}, while the creation and the application of the coarse-grid Dslash operator are implemented in \texttt{CoarsenedMatrix}.
Since Grid places particular emphasis on generic code we can reuse these classes for implementing an MG solver for a different fermion formulation.\footnote{Likewise, all performance improvements achieved within theses classes will benefit other fermion formulations.} 
The methods mentioned above are sufficient for a two-level MG method.
However, the need for larger and finer lattices with physical quark masses requires the addition of more levels.
To be able to support an arbitrary number of levels we enable the coarse matrix to recursively be projected to a coarser grid.
Adhering to the coarsening interface, this only requires us to implement two member functions of \texttt{CoarsenedMatrix}, i.e., \texttt{Mdiag}, which applies the diagonal part of the Dirac matrix to a vector, as well as \texttt{Mdir}, which applies the hopping term in a single direction.

On top of these classes we implement the actual solver class and thus the solver's interface to the outside world.
Reflecting the recursive nature of the MG algorithm, the solver class recursively calls instances of itself using templating of the different matrix types.

The basic setup used in HDCR, which we will refer to as initial setup from now on, iterates a smoother on random vectors to find near-null vectors, i.e., approximate low modes of the Dirac operator, and then constructs the relevant multigrid operators.
At the start of our work, this functionality was already present in \texttt{Aggregation}.
Our implementation adds the more sophisticated setup proposed by \DDaAMG: After running the initial setup based on the smoother, it utilizes the entire MG preconditioner to perform an iterative refinement procedure for a better approximation of the near-null space.
More setup iterations significantly accelerate the solver but imply additional numerical cost.
Therefore, there is a tradeoff between the number of setup iterations and the number of solves a particular setup can be used for \cite{Osborn:2010mb}.
The optimal number of setup iterations will be different for HMC and analysis runs.

Multigrid for Wilson-like fermions is run directly on the Dirac operator $D$, not on the squared operator $D^{\dagger}D$.
Hence solvers for non-Hermitian matrices are used within the multigrid method.
Standard choices here are algorithms based on conjugate and minimal residuals.
We introduce variants of the latter type within the Grid framework: MinimalResidual (MR), its generalized form (GMRES), and the corresponding flexible version (FGMRES).
The latter allows for right-preconditioning by a non-stationary preconditioner matrix and is thus well suited as an outer Krylov solver.
At the time of the conference we had fixed the smoother on every MG level and the coarse-grid solver to GMRES.

Normally the multigrid preconditioner is run in lower precision than the outer solver.
Grid's \texttt{precisionChange} enables our implementation to go from an MG solver fixed to one precision (double or single) to a mixed-precision approach (outer solver in double, preconditioner in single) by changing one line of code.

At the time of the conference, our implementation was still missing some attractive features typically used in MG inverters, e.g., the Schwarz Alternating Procedure (SAP), which is commonly used as a smoother on all MG levels, or red-black preconditioning.
In the meantime we have been working on including these and other features \cite{Richtmann:github}.

\subsection{Chirality}
\label{sec:chirality}

A successful multigrid for Wilson clover fermions must preserve chirality through the coarsening of the Dirac operator, i.e., left- and right-handed components must not be mixed.
This way, the $\gamma_5$-Hermiticity of the Dirac operator acting on the spin d.o.f.\ translates to $\sigma_3$-Hermiticity acting on the coarse chirality d.o.f.
Usually this is realized by separating the left- and right-handed components of the null-space vectors when creating the intergrid operators from them, i.e., the prolongation operator reads
$P = \left[ \frac{1}{2} (1+\gamma_5) \psi_i, \frac{1}{2} (1-\gamma_5) \psi_i \right]$.
Due to limitations in Grid's coarsening infrastructure we are forced to store the null-space vectors in a chirally-doubled format.
Obviously, this wastes memory (as explicit zeros are stored), doubles the memory requirement on the null-space vectors, and leads to unnecessary work to be performed in the setup.
Again, this is something we have been improving on since the time of the conference.

\section{Performance tuning}

\begin{figure}[ht]
    \centering
    \includegraphics[]{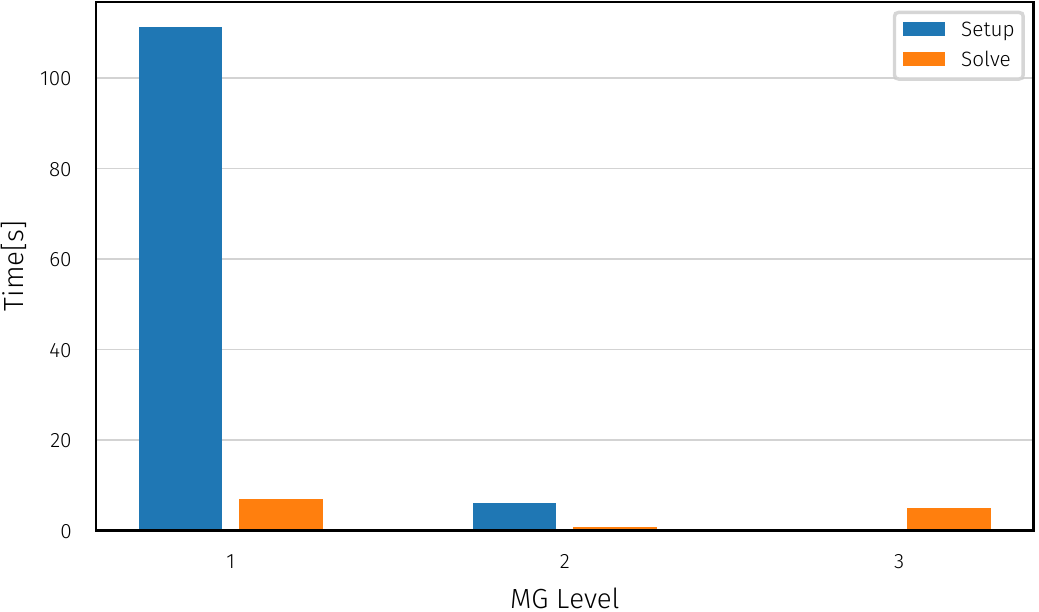}
    \caption{\label{fig:overview_total_runtime_barplot} Run-time distribution of the 3-level MG method with default \DDaAMG parameters on a $16^4$ lattice, run on a single Xeon Phi 7210 (KNL) node. Level 1 is the finest level.}
\end{figure}

\begin{figure}[b]
    \centering
    \includegraphics[]{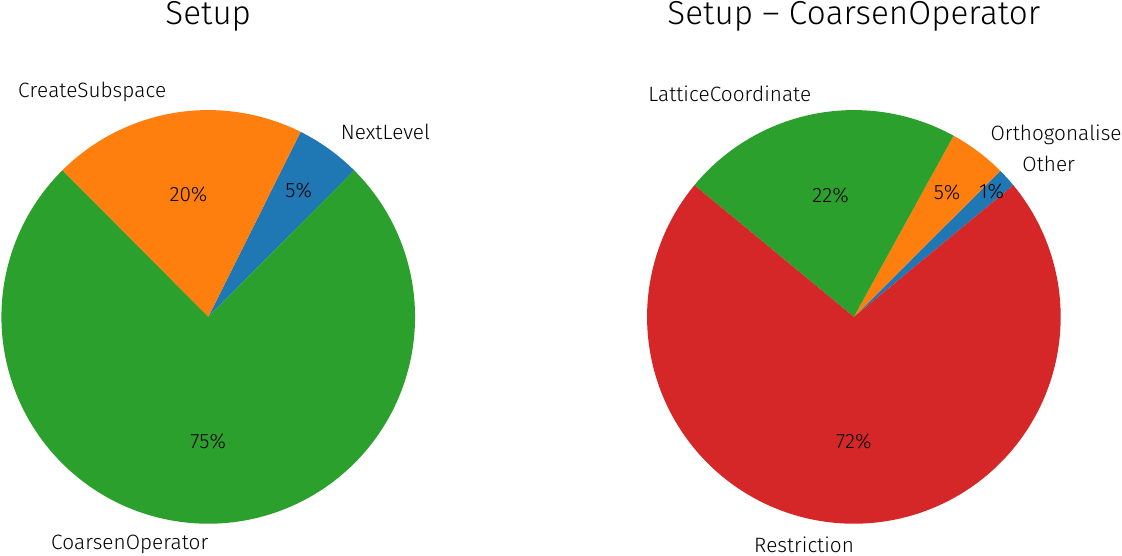}
    \caption{\label{fig:overview_setup_runtime_level_0_piechart} Left: Run-time contribution of the setup in Figure~\ref{fig:overview_total_runtime_barplot}. The green and orange parts correspond to the finest level, while the blue part contains all other levels. Right: Run-time contribution of the operator coarsening on the finest level.}
\end{figure}

\begin{figure}[ht]
    \centering
    \includegraphics[]{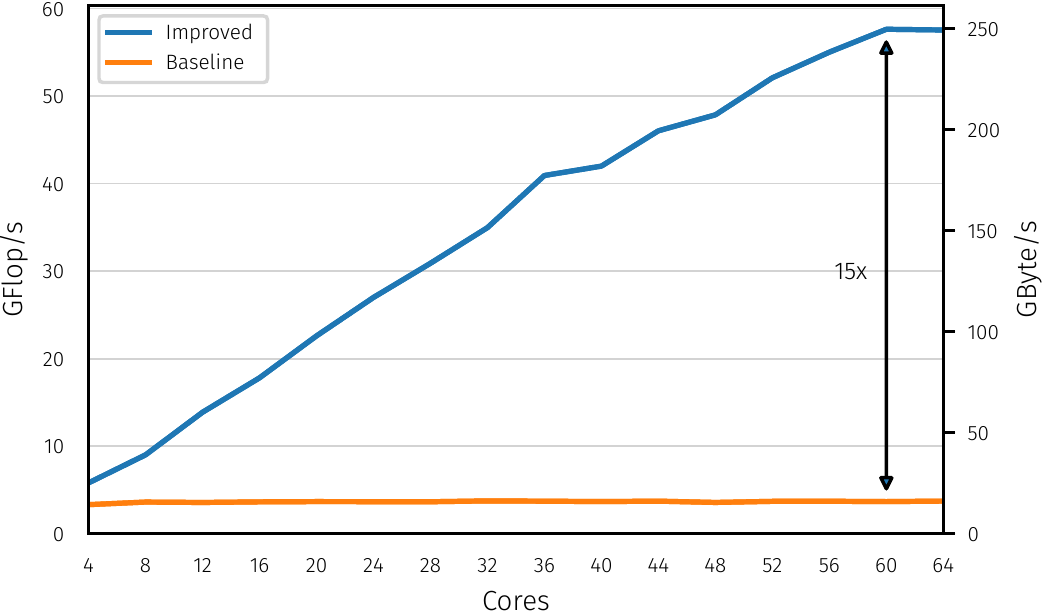}
    \caption{\label{fig:restriction_performance_comparison} \texttt{blockProject} threaded over fine (blue) and coarse (orange) lattice sites.}
\end{figure}

\begin{figure}[ht]
    \centering
    \includegraphics[]{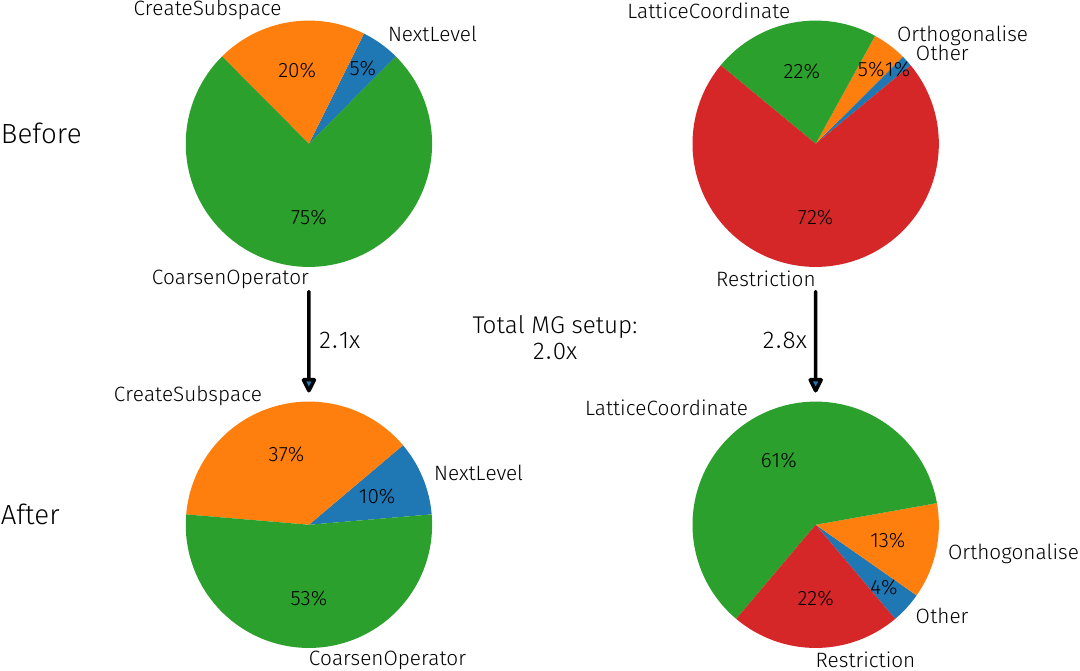}
    \caption{\label{fig:overview_setup_runtime_level_0_after_optimization_piechart} Performance improvements resulting from our changes to \texttt{blockProject}.}
\end{figure}

After implementing an almost feature-complete MG algorithm we now consider initial performance characteristics of our code.
We first run a small $16^4$ test lattice on a single node with the default parameter set of \DDaAMG and investigate the run-time distribution.
As Figure~\ref{fig:overview_total_runtime_barplot} shows, most of the run time is spent in the setup on the finest level.
Therefore we show a more detailed look at its run-time contributions in Figure~\ref{fig:overview_setup_runtime_level_0_piechart}.
The construction of the coarse link matrices (\texttt{CoarsenOperator}) dominates the run time, the major contribution being the application of the restriction operator.
Taking a close look at this function (\texttt{blockProject}) we quickly discover the reason.
The code is threaded over the sites of the fine lattice, which requires a critical region when writing to the corresponding coarse site in order to avoid race conditions.
Therefore only one thread runs at a given time, which significantly deteriorates performance.
We alleviate this problem by threading over the sites of the coarse lattice.
Although this requires the introduction of a lookup table (calculated serially) it removes the potential race condition since each thread only writes to its set of coarse sites and hence enables the function to be run in parallel.
Figure~\ref{fig:restriction_performance_comparison} illustrates the performance improvement we achieve with this change.
We are able to speed up this function by a factor of 15 and sustain a memory bandwidth of 250~GB/s.
The KNL we are using has a peak memory bandwidth of $\sim\,$440~GB/s from high-bandwidth MCDRAM.
The reason why we are not able to saturate the memory bandwidth lies in the memory access pattern on the fine lattice we introduced with our changes.
We verified this by running a benchmark with a linearized lookup table.
Doing so we are able to sustain the wire bandwidth.
With our changes to \texttt{blockProject} we were able to reduce the total run-time contribution of this function from 54\% to 11\% so that it no longer is the dominating contribution.
From a practical point of view, it is not worth putting in further optimization efforts until it becomes the bottleneck again.
In total our modification accelerated the entire MG setup by a factor of 2, as depicted in Figure~\ref{fig:overview_setup_runtime_level_0_after_optimization_piechart}.

\section{A first comparison with \DDaAMG}

We now compare the present state of our MG implementation to the current go-to MG library for CPUs, DDalphaAMG \cite{Rottmann:github},\footnote{See \cite{Heybrock:2015kpy,Georg:2017zua,Ishikawa:2018fad} for ports of \DDaAMG to the Intel Xeon Phi (KNC, KNL) and the K computer.} to see how we compare in terms of solver wall-clock time.
We do this comparison on a standard Intel Xeon CPU of the Broadwell series, as both code bases have support for it.
Note that \DDaAMG normally uses SAP as its smoother.
However, we choose to run it with GMRES to get a fair comparison, as our code does not feature SAP yet. 
Apart from that we let both code bases use their full set of features, i.e., the most modern ISA they support (SSE for \DDaAMG and AVX2 for Grid). For \DDaAMG we enable even-odd preconditioning, which our implementation does not feature yet.
Table~\ref{tab:comparison_parameters} summarizes the parameters of the benchmark run.
Solver parameters not explicitly stated are set to their best values proposed by the authors of \DDaAMG \cite{Frommer:2013fsa}.

\begin{table}[t]
    \centering
        \begin{tabular}{cccccc}
          \toprule
          System           & Lattice & Blocksize & Mass    & Basis vectors & Smoother \\
          \midrule
          $8$-core Broadwell & $16^4$   & $4^4$     & $-0.25$ & $20$          & GMRES \\
          \bottomrule
        \end{tabular}
    \caption{Parameters for the comparison of \DDaAMG and our current Grid MG implementation.}
    \label{tab:comparison_parameters}
\end{table}

In Table~\ref{tab:comparison_results} we show the run times of the various algorithmic components as a function of the number of setup iterations.
Our Grid MG implementation is consistent with \DDaAMG's behavior in terms of outer solver iteration counts.
This gives us further confidence (in addition to internal checks in the code) that our method was implemented correctly.
We observe that, compared to \DDaAMG, the setup phase of Grid MG takes about twice longer, while the solve is about twice faster.
Grid MG does worse in the setup since (i) due to the chiral doubling, see Sec.~\ref{sec:chirality}, the number of null-space vectors is twice that of \DDaAMG and (ii) it does more work than \DDaAMG (9 vs 5 stencil points in \texttt{CoarsenOperator}).
These effects are alleviated by the length of the SIMD vectors (128 bits for SSE, 256 bits for AVX2).
The SIMD length also explains the faster solve of Grid MG, which is not affected by (i) and (ii).

\begin{table}[b]
    \centering
        \begin{tabular}{c|@{\hspace*{3mm}}cc|rr|rr|rr|rr}
          \toprule
          & \multicolumn{2}{@{\hspace*{-1mm}}c|}{Outer Iter.} & \multicolumn{2}{c|}{Initial Setup} & \multicolumn{2}{c|}{Iterative Setup} & \multicolumn{2}{c|}{Solve} & \multicolumn{2}{c}{Total} \\
          & & & & & & & & & & \\[-4.5mm]
          \specialrule{.4pt}{0pt}{0pt}
          \diagbox[width=26mm]{\hspace*{-2mm}Setup iter.}{\raisebox{-1mm}{Code}} & D &  G & \multicolumn1cD & \multicolumn1{c|}G &    \multicolumn1cD & \multicolumn1{c|}G & \multicolumn1cD & \multicolumn1{c|}G & \multicolumn1cD & \multicolumn1cG \\
          \specialrule{.4pt}{0pt}{0pt}
          & & & & & & & & & & \\[-4mm]
          0  & 98 & 98 & 5.48 & 16.9  & \hspace{5.5mm} 0    & 0     \hspace{0mm} & 19.4 & 11.7 & 24.9 & 28.6  \\
          1  & 79 & 78 & 5.60 & 16.2  &                5.0  & 13.4  \hspace{0mm} & 16.5 & 9.49 & 27.1 & 39.1  \\
          2  & 48 & 59 & 5.56 & 16.3  &                10.1 & 26.4  \hspace{0mm} & 11.4 & 7.52 & 27.1 & 50.2  \\
          3  & 31 & 42 & 5.67 & 16.4  &                16.4 & 38.9  \hspace{0mm} & 8.34 & 5.88 & 30.4 & 61.2  \\
          4  & 26 & 32 & 5.53 & 16.5  &                23.5 & 52.5  \hspace{0mm} & 7.04 & 4.63 & 36.1 & 73.6  \\
          5  & 25 & 27 & 5.60 & 16.5  &                30.1 & 65.6  \hspace{0mm} & 6.77 & 3.96 & 42.5 & 86.1  \\
          6  & 25 & 26 & 5.50 & 16.5  &                35.8 & 78.6  \hspace{0mm} & 6.78 & 3.80 & 48.1 & 98.9  \\
          7  & 25 & 25 & 5.51 & 16.1  &                43.0 & 92.3  \hspace{0mm} & 6.82 & 3.70 & 55.3 & 112.1 \\
          8  & 25 & 25 & 5.69 & 16.2  &                49.1 & 104.9 \hspace{0mm} & 6.86 & 3.69 & 61.6 & 124.8 \\
          9  & 25 & 25 & 5.71 & 16.6  &                55.4 & 118.4 \hspace{0mm} & 6.82 & 3.72 & 67.9 & 138.7 \\
          10 & 26 & 25 & 5.66 & 16.2  &                62.0 & 131.9 \hspace{0mm} & 7.12 & 3.72 & 74.8 & 151.8 \\
          \bottomrule
        \end{tabular}
        \caption{Run-time comparison of \DDaAMG (D) and Grid MG (G). For details see text.}
    \label{tab:comparison_results}
\end{table}

Note that Table~\ref{tab:comparison_results} is a snapshot corresponding to the results presented at the conference. In the meantime, the improvements mentioned in Section~\ref{sec:implementation} led to a total run time of our Grid MG that is about 60\% that of \DDaAMG on the same lattice.

\section{Summary and outlook}

We implemented a multigrid solver for Wilson clover fermions within the Grid framework and started to improve its performance.
While our implementation passes algorithmic regression against the \DDaAMG library, at the time of the conference the time to solution of our code was higher than that of \DDaAMG. 
In the meantime we have improved the performance of our code and implemented missing parts of the algorithm \cite{Richtmann:github}. 

\section*{Acknowledgment}

We acknowledge funding provided by the German Research Foundation (DFG) in the framework of SFB/TRR-55.

\bibliographystyle{JHEP_TW}
\bibliography{references}

\end{document}